\documentclass[aps,prl,twocolumn,showpacs,superscriptaddress,nofootinbib]{revtex4-1}  

\usepackage{graphicx}  
\usepackage{dcolumn}   
\usepackage{amssymb}   
\usepackage[caption=false]{subfig}
\usepackage{amsmath}
\usepackage{siunitx}

\begin{document}

\title{Multi-mode double-bright EIT cooling}
\author{Nils Scharnhorst}
	\affiliation{QUEST Institute for Experimental Quantum Metrology, Physikalisch-Technische Bundesanstalt, 38116 Braunschweig,~Germany}
	\affiliation{Institut f{\"u}r Quantenoptik, Leibniz Universit{\"a}t Hannover, 30167 Hannover,~Germany}
\author{Javier Cerrillo}     
	\affiliation{Faculty II Mathematics and Natural Sciences, Institute for Theoretical Physics, Technische Universit\"at Berlin, 10623 Berlin,~Germany}
\author{Johannes Kramer}
	\affiliation{QUEST Institute for Experimental Quantum Metrology, Physikalisch-Technische Bundesanstalt, 38116 Braunschweig,~Germany}
\author{Ian D. Leroux}
	\altaffiliation[Current adrdress: ]{National Research Council Canada, Ottawa, Ontario, K1A 0R6, Canada}
	\affiliation{QUEST Institute for Experimental Quantum Metrology, Physikalisch-Technische Bundesanstalt, 38116 Braunschweig,~Germany}
\author{Jannes B. W{\"u}bbena}
	\altaffiliation[Current adrdress: ]{Geo++ GmbH, 30827 Garbsen}
	\affiliation{QUEST Institute for Experimental Quantum Metrology, Physikalisch-Technische Bundesanstalt, 38116 Braunschweig,~Germany}
\author{Alex Retzker}
	\affiliation{Racah Institute of Physics, Hebrew University of Jerusalem, 91904 
	Jerusalem, Israel}
\author{Piet O. Schmidt}
	\affiliation{QUEST Institute for Experimental Quantum Metrology, Physikalisch-Technische Bundesanstalt, 38116 Braunschweig,~Germany}
	\affiliation{Institut f{\"u}r Quantenoptik, Leibniz Universit{\"a}t Hannover, 30167 Hannover,~Germany}
\date{\today}

\begin{abstract}
We developed a multi-mode ground state cooling technique based on electromagnetically-induced transparency (EIT). By involving an additional ground and excited state, two individually adjustable bright states together with a dark state are created. While the dark state suppresses carrier scattering, the two bright states are brought into resonance with spectrally separated motional red sidebands. With this double-bright EIT (D-EIT) cooling scheme, we experimentally demonstrate ground state cooling of all three motional modes of a trapped $^{40}$Ca$^+$ with a single cooling pulse. The approach is scalable to more than two bright states by introducing laser couplings to additional states. 
\end{abstract}

\pacs{42.50.-p, 03.75.Be, 37.10.De, 37.10.Mn, 06.30.Ft}
\maketitle

Ground state cooling (GSC) of trapped ions and neutral atoms is a necessity in many quantum optics experiments such as quantum simulations \cite{porras_effective_2004, blatt_quantum_2012},  quantum state engineering \cite{monz_14-qubit_2011, leibfried_creation_2005, haffner_scalable_2005, johnson_ultrafast_2016, lo_spin-motion_2015}, quantum logic spectroscopy \cite{wan_precision_2014, wolf_non-destructive_2016, rosenband_frrequency_2008, Schmidt2005, hempel_entanglement-enhanced_2013, chou_preparation_2017}, and quantum logic clocks \cite{rosenband_frrequency_2008, Ludlow_ClocksReview} in order to minimize motion-induced errors and to obtain full quantum control over the system. Cooling performance is characterized by the cooling rate, the minimal achievable kinetic energy, and how many motional modes are cooled simultaneously. Fast cooling and thus short cooling times are important in all applications to reduce the associated dead time. This is particularly important in optical clocks where dead time directly leads to reduced stability through the Dick effect \cite{Ludlow_ClocksReview, poli_optical_2013, dick_local_1987}, limiting applications e.g. in relativistic geodesy \cite{vermeer_chronometric_1983, bjerhammar_relativistic_1985, lisdat_clock_2016}, or tests of fundamental physics \cite{rosenband_frrequency_2008, godun_frequency_2014, huntemann_improved_2014}. Standard techniques for GSC include sideband cooling \cite{diedrich_laser_1989, monroe_resolved-sideband_1995, roos_quantum_1999, vuletic_degenerate_1998, han_3d_2000, hamann_resolved-sideband_1998} and more recently, fast cooling via electromagnetically induced transparency (EIT) \cite{Morigi2000, roos_experimental_2000, lin_sympathetic_2013, kampschulte_eit-control_2012, Lechner2016_EIT}.  Sideband cooling is implemented in the resolved sideband regime and thus capable of cooling only within a narrow spectral band typically containing only one motional mode. Compared to this, EIT cooling has the advantage that the asymmetry of the Fano-like scattering profile can be employed to cool several modes over a larger frequency range in a single pulse \cite{Lechner2016_EIT}. However, for mode frequencies that are separated by more than the width of the Fano profile, cooling becomes less efficient, and some modes may even be heated. 

Here, we demonstrate a novel scalable approach to standard EIT cooling which is hereafter referred to as double-bright EIT (D-EIT) cooling. In this scheme a third ground state is coupled to a second excited state via an additional laser. The additional degrees of freedom can be used either to suppress a higher order heating process \cite{evers_double-eit_2004} or to cool a second band of mode frequencies with the same cooling pulse. We experimentally demonstrate multi-mode ground state cooling of all degrees of freedom of a single $^{40}$Ca$^+$ ion via D-EIT cooling in a single cooling pulse. The experimental results on cooling rates and final temperatures are supported by full density matrix calculations. 

\begin{figure}
	\includegraphics[width=.99\linewidth]{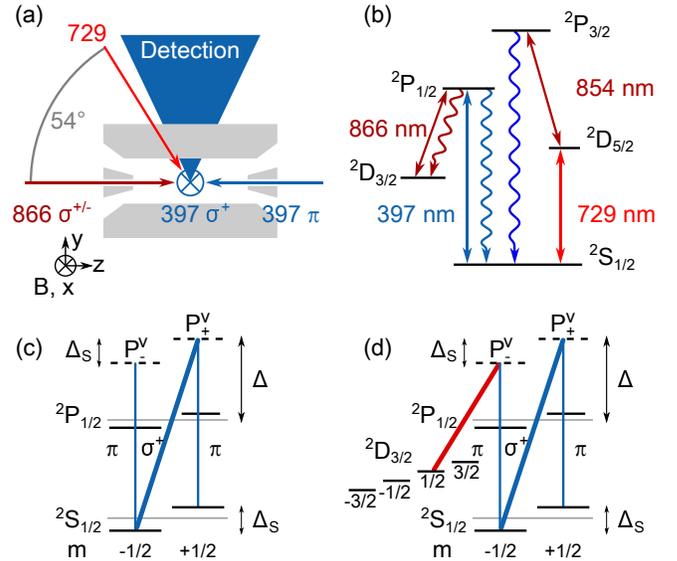}
	\caption{\label{levelslayout}(color online) (a) Schematic of the beam setup with respect to the trap and magnetic field $B$. (b) Energy levels of $^{40}$Ca$^+$. (c) Relevant levels for single EIT cooling. (d) Relevant levels for D-EIT. In (c), (d) the pump beam(s) are drawn with thick lines.}
\end{figure}

A single $^{40}$Ca$^+$ ion is stored in a linear Innsbruck-design Paul trap as described in \cite{Dolezal2015_BBR, Wuebbena2014, Scharnhorst_PRA}, having radial vibrational mode frequencies $\nu_r$ of \SI{2.552}{\mega\hertz} and \SI{2.540}{\mega\hertz}, and an axial vibrational mode frequency $\nu_a$ of \SI{904.6}{\kilo\hertz}. Three pairs of field coils create a constant magnetic field of \SI{416}{\micro\tesla} perpendicular to the trap axis (see Fig.~\ref{levelslayout}). For further experimental details see \cite{Scharnhorst_PRA}.

Fig.~\ref{levelslayout}(b) shows the relevant energy levels of $^{40}$Ca$^+$. All measurements use the same experimental pulse order. First, the ion is Doppler cooled in all directions for 1~ms using the two laser beams at \SI{397}{\nano\meter} shown in Fig.~\ref{levelslayout}(a). One of them cools only the radial degrees of freedom and is $\sigma^+$ polarized, the other one is directed along the trap axis and is $\pi$ polarized. A counter-propagating \SI{866}{\nano\meter} laser acts as a repumper of the $D_{3/2}$ state during Doppler cooling and is $\sigma^{\pm}$ polarized. A D-EIT cooling pulse of length $t_\textrm{c}$ is followed by a low-power \SI{2}{\micro\second} optical pumping pulse of \SI{397}{\nano\meter} $\sigma^+$ and \SI{866}{\nano\meter} light. This prepares the $|^2S_{1/2},m_\mathrm{J}=+1/2\rangle$ state with high fidelity and negligible ($\Delta\bar{n}\ll0.04$) motional heating \cite{Scharnhorst_PRA}. Motional state detection is performed via the electron shelving technique \cite{Dehmelt1975_shelving, roos_quantum_1999} on motional sidebands of the $^2S_{1/2}$ to $^2D_\mathrm{5/2}$ transition. During a final \SI{397}{\nano\meter} $\pi$-polarized and \SI{866}{\nano\meter} detection pulse of \SI{250}{\micro\second} duration, fluorescence is detected with high-numerical aperture imaging optics and a photo-multiplier tube. The \SI{729}{\nano\meter} state detection (logic) laser beam has a projection onto the axial and radial directions as shown in Fig.~\ref{levelslayout}(a). For each setting the excitation probability is determined by averaging over 250 repetitions. The \SI{729}{\nano\meter}, \SI{866}{\nano\meter}, and \SI{397}{\nano\meter} lasers are phase locked to the same ultra-stable reference using the high-bandwidth transfer lock technique described in \cite{Scharnhorst2015_lock}. This maintains a definite phase relationship among he cooling lasers, as demonstrated in \cite{Scharnhorst_PRA}.

For a single trapped particle a mean motional quantum number can be assigned to each of its motional degrees of freedom (two radial modes, one axial mode). The temporal evolution of the mean occupation, $\bar{n}(t)$, and its steady-state value $n_\mathrm{ss}$ are determined using sideband thermometry on the logic transition \cite{turchette_nbar_2000} with pulse lengths optimizing the signal-to-noise ratio \cite{Scharnhorst_PRA}. The resulting $\bar{n}(t)$ is fitted with an exponential decay whose decay constant $R$ is defined as the cooling rate. As shown in Fig.~\ref{levelslayout}(a) all cooling beams run in the same plane to which the logic laser has an angle $\alpha=(54.7\pm5.1)^{\circ}$. The two orthogonal radial modes are rotated around the axial direction ($z$) such that the modes have a $(26\pm10)^{\circ}$ and $(-64\pm10)^{\circ}$ angle with respect to the $xz$-plane of the cooling lasers. In radial direction, $\bar{n}$ is obtained from the mode that has a higher projection onto the logic laser and smaller projection onto the cooling lasers. Since the other radial mode has a higher projection onto the plane spanned by the cooling beams, we expect it to be cooled even more efficiently.

\begin{figure}
	\includegraphics[width=.99\linewidth]{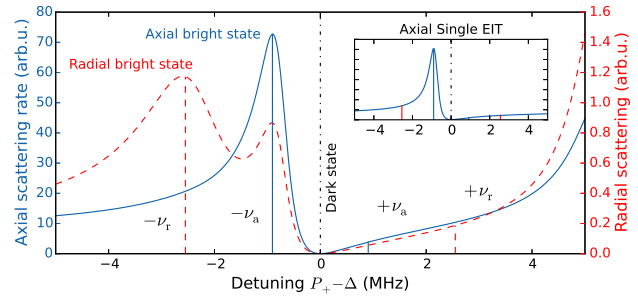}
	\caption{\label{EITscat}(color online) Simulated scattering rate during D-EIT cooling versus detuning from the virtual level $P_+$. Vertical straight lines indicate motional frequencies. The inset shows single EIT spectrum in the respective direction optimized for axial cooling. D-EIT has a bright state at both radial and axial frequency, while for single EIT there is only one bright state. The position of the radial (axial) bright state is dominated by the Rabi frequency of the \SI{397}{\nano\meter} $\sigma^+$ (\SI{866}{\nano\meter} $\sigma^-$) beam.}
\end{figure}
EIT cooling is based on the coherent engineering of dark states in a $\Lambda$-like level scheme with two ground states and one shared excited state \cite{Morigi2000}. Fig.~\ref{levelslayout}(c) shows the relevant levels and beams for single EIT cooling in $^{40}$Ca$^+$. Here, the ground states are the $|^2S_{1/2}, m_\mathrm{J}=\pm1/2\rangle$ levels and the excited state is chosen to be the $|^2P_{1/2}, m_\mathrm{J}=1/2\rangle$ state. If lasers connecting the respective ground states are brought into two-photon resonance near the excited state, a dark resonance is generated, as shown in the simulated scattering spectrum in Fig.~\ref{EITscat}. Each dark resonance is associated with a bright state that is theoretically described by a Fano-profile \cite{Janik_Fano}. Its spectral position (shift) with respect to the dark resonance depends on the intensity of the two lasers forming the dark resonance. Typically, one of the two lasers (pump) is stronger than the other one (probe) such that the shift of the bright state is dominated by the pump beam. In our setup the \SI{397}{\nano\meter} $\sigma^+$ beam is used as pump and the \SI{397}{\nano\meter} $\pi$ beam serves as probe beam. Pump and probe are brought into resonance on a virtual energy level $P_\textrm{+}^v$ which is blue detuned by $\Delta$ with respect to the unperturbed $^2P_{1/2}$ level. 
For a Rabi frequency of the probe
laser $\Omega_{\pi}$, a total Lamb-Dicke parameter projection on
the relevant axis $\eta$, and a decay rate of the excited state $\gamma$,
an optimal cooling rate $R^{EIT}\simeq\frac{\eta^{2}\Omega_{\pi}^{2}}{2\gamma}$
towards a final occupation number $n_{\mathrm{ss}}^{EIT}\simeq\frac{\gamma^{2}}{4\Delta^{2}+\gamma^{2}}\left(1+2\frac{\Omega_{\pi}^{2}}{\Omega^{2}}\right)$
can be achieved, where $\Omega\equiv\sqrt{\Omega_{\pi}^2 + \Omega_{\sigma}^2}$, and the term proportional to $\Omega_{\pi}^2$ corresponds
to a heating contribution from spurious $\pi$ coupling to $|^2P_{1/2}, m_\mathrm{J}=-1/2\rangle$. In
order to curb this contribution, a reduced $\Omega_{\pi}$ needs to
be chosen, which in practice constraints the ratio between the cooling
rate and the trap frequency to values much smaller than the maximum
$\frac{R^{EIT}}{\nu_{a,r}}\ll\frac{2\Delta}{\gamma}\eta_\mathrm{a,r}^{2}$ achievable
with EIT \cite{Scharnhorst_PRA}.

In D-EIT cooling, the $\Lambda$ scheme of single EIT cooling is extended to a double $\Lambda$-shaped level scheme, as shown in Fig.~\ref{levelslayout}(d). This is accomplished by introducing an additional coupling between the $|^2D_{3/2}, m_\mathrm{J}=1/2\rangle=|D_+\rangle$ and virtual $P_-^v$ level through an \SI{866}{\nano\meter} beam acting as an additional pump. It creates a second bright state together with the $|S_-\rangle\leftrightarrow P_-^v$ $\pi$ coupling. For our beam geometry, it is formed by purely axial beams and the additional bright state can be brought into resonance with the axial motional sideband by an appropriate choice of the \SI{866}{\nano\meter} pump beam Rabi frequency. Similarly, the bright state formed by \SI{397}{\nano\meter} $\pi$ and $\sigma^+$ couplings is shifted into resonance with the radial sidebands by choosing the corresponding pump beam Rabi frequency $\Omega_{\sigma^+}$. This way, both radial and axial modes that are spectrally separated in our case by more than \SI{1.5}{\mega\hertz} can be cooled simultaneously. Furthermore, the off-resonant excitation of the $|^2P_{1/2}, m_\mathrm{J}=-1/2\rangle$ level is suppressed via coupling of this level to the $^2$D$_{3/2}$-manifold. 
This eliminates the contribution proportional
to $(\Omega_{\pi})^2$ in the final occupation, so that $n_{\mathrm{ss}}^{D-EIT}\simeq\frac{\gamma^{2}}{4\Delta^{2}+\gamma^{2}}$
and the ratio $\frac{R_{r,a}^{D-EIT}}{\nu_{a,r}}\simeq\frac{2\Delta}{\text{\ensuremath{\gamma}}}\eta_{\sigma,\pi}^{2}$
can now approach its maximum value \cite{Scharnhorst_PRA}. 

However, these limits are obtained from a perturbative approach using the Lamb-Dicke approximation, which requires $\eta_{i}\Omega_{i}\ll\nu_{a,r}$
for all laser couplings $i$, so that they become unreliable for large
Rabi frequencies. For a cooling rate of the form $R_{a,r}\simeq\frac{\eta_{i}^{2}\Omega_{j}^{2}}{2\gamma}$,
this implies that the prediction is only valid as long as $R_{a,r}\ll\frac{\nu_{a,r}^{2}}{2\gamma}$,
which in our setting translates to a cooling rate much lower than $10^{5}\:\mathrm{phonons}/\mathrm{s}$
for the axial degree of freedom and $10^{6}\:\mathrm{phonons}/\mathrm{s}$
for the radial degrees of freedom. Since the Stark shift must be maintained to keep the bright states on resonance with the motional modes, these conditions can only be satisfied by increasing the detuning $\Delta$. 

\begin{figure}
	\includegraphics[width=1\columnwidth]{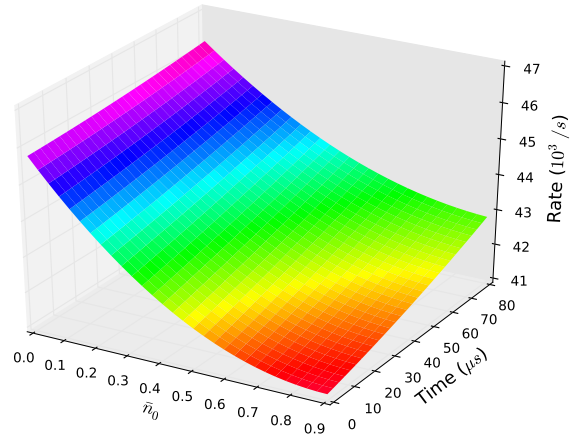}\protect\caption{\label{fig:Time-dependent-rate}Time dependent rate $R\left(t, \bar{n}_0\right)$
		in units of $10^{3}$ phonon$\cdot s^{-1}$ for axial single EIT at $\Delta=3\Gamma$ as a function of the initial occupation number $n_{0}$ and the cooling time. The Lamb-Dicke prediction for the rate corresponds to $52.2\times10^{3}$ phonon$\cdot s^{-1}$, whereas the measured value is $38.2\times10^{3}$ phonon$\cdot s^{-1}$.}
\end{figure}

Beyond the perturbative limit, the occupation number does not follow an exponential decay during cooling, but it is possible to define a decay rate with the help of a generalized fluctuation-dissipation theorem \cite{Cerrillo2016}.  The effective decay rate depends on the initial oscillator temperature $\bar{n}_0$ and varies as the steady state is approached \cite{Scharnhorst_PRA}. In practical terms, one observes that the time-dependent rate is lower than the Lamb-Dicke prediction. To theoretically evaluate the D-EIT cooling process outside the Lamb-Dicke regime, we performed numerical full density matrix master equation simulations, taking into account all 8 relevant electronic and the lowest 17 motional states \cite{Scharnhorst_PRA}. An instance of this is presented in Fig.~\ref{fig:Time-dependent-rate}, where $R(\bar{n}_0, t)$ is shown as a function of the elapsed cooling time $t$ and initial temperature $\bar{n}_0$. $R(\bar{n}_0, t)$ falls as $\bar{n}_0$ increases, but it recovers as a function of time as lower thermal states become involved in the dynamics.  
This effect is connected with the observation of an increase of the $^2$P$_{1/2}$-state population when leaving the Lamb-Dicke regime. A possible explanation for this observation is that the strong laser-induced spin-motional coupling prevents establishing an electronic dark state that has no admixture of the excited $^2$P$_{1/2}$-state. Increasing $^2$P$_{1/2}$ population results in an enhanced scattering on carrier and sidebands, limiting the cooling rate and achievable $n_\mathrm{ss}$. The effect is more pronounced the higher Fock states are involved, but even at an $\bar{n}<1$ a \SI{20}{\%} reduction in cooling rate can be observed (see Fig.~\ref{fig:Time-dependent-rate}). This behavior is characteristic of strong coupling dissipative theory \cite{Cerrillo2016}. A full understanding of this behavior requires the development of an analytical model for cooling in the regime beyond the LD approximation, which is subject to future theoretical investigations.

\begin{figure}
	\includegraphics[width=1.\linewidth]{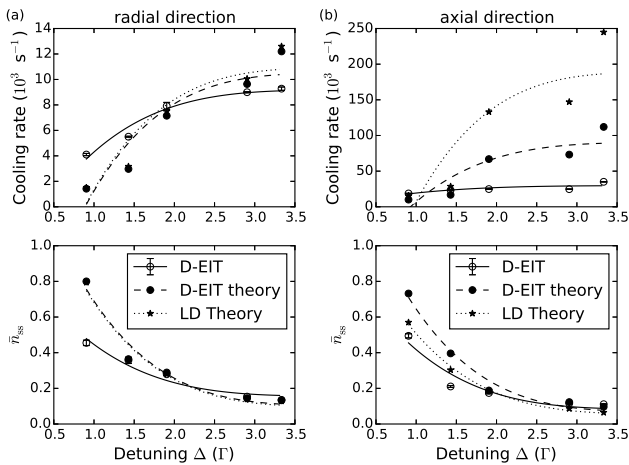}		
	\caption{Cooling rates and equilibrium $\bar{n}$ for D-EIT as a function of the blue detuning $\Delta$ of the virtual level $P_+$. (a) Radial modes. (b) Axial mode. In each case, the experimental values (empty circles), simulated values from the full master equation theory (solid cirles), and simulated values from the Lamb-Dicke theory (stars) are given. The solid, dashed, and dotted lines are quartic functions fitted to the data points to guide the eye.\label{expresults}}
\end{figure}

In Fig.~\ref{expresults} we compare the experimental results of the cooling rates and $n_\mathrm{ss}$ with simulated data from Lamb theory, and full density matrix theory as described in \cite{Scharnhorst_PRA}. For each data point the experimentally obtained $\Omega_\mathrm{\pi}$ and $\Delta$ are used in the simulation and the Rabi frequencies of the pump beams are chosen such that the corresponding bright states coincide with the radial and axial mode frequencies. The given simulated $n_\mathrm{ss}$ and cooling rates are derived via sideband thermometry as in the experimental case. To compare a simulated time-depended cooling rate to the experimental data, the simulated rate is given at $\bar{n}(t)=1$. 

Fig.~\ref{expresults} displays the cooling rates and $n_\mathrm{ss}$ for the radial and axial modes for D-EIT cooling, the Lamb-Dicke simulations, and the full master equation simulations, as a function of the blue detuning $\Delta$ of the virtual $P_\textrm{+}^v$ from the magnetic field-free $^2P_{1/2}$ level in units of the natural linewidth of $\Gamma=2\pi\times$\SI{20.7}{MHz} of the $^2$S$_{1/2}$ to $^2$P$_{1/2}$ transition. The radial and axial modes are cooled simultaneously within the same D-EIT cooling pulse. Increasing detuning yields both a colder steady state and faster cooling. The optimal cooling parameters were found experimentally by adjusting the three involved Rabi frequencies to obtain ac Stark shifts of the bright states matching the two mode frequencies, while maximizing the cooling rate and minimizing the steady-state $n_\mathrm{ss}$ \cite{Scharnhorst_PRA}. Therefore, larger detunings allow higher Rabi frequencies and thus stronger cooling rates. As the detuning is increased, the bright state scattering resonances narrow  while the residual blue sideband excitations are suppressed. This results in an improved ratio between cooling and heating scattering events and thus a lower final $n_\mathrm{ss}$. Initially starting from Doppler cooling temperatures of $\bar{n}\sim11.1$ in axial and $\bar{n}\sim3.6$ in radial direction, D-EIT cooling allows to reach $n_\textrm{ss,a}=0.11$ and $n_\textrm{ss,r}=0.14$ in a single cooling pulse of \SI{670}{\micro\second} length, for $\Delta=3.4\:\Gamma$.
 
In the radial direction, the experimental $n_\mathrm{ss}$ and the cooling rates agree satisfactorily with full density matrix and Lamb-Dicke simulations. In the axial direction a clear deviation of the measured rates from the Lamb-Dicke theory is observed for large cooling rates. However, even the full density matrix simulations predict larger cooling rates than experimentally observed. This difference increases with $\Delta$ and may be explained by the limited maximum number of $17$ Fock states that can be numerically simulated. The truncation of high Fock states in the simulation with non-negligible occupation leads to an overestimation of $R_\mathrm{a}$ and it is verified, that $R_\mathrm{a}$ decreases with increasing number of simulated Fock states. This effect is more pronounced in the axial direction where the lower trap frequency results in population of larger Fock states compared to the radial direction for the same initial temperature. Furthermore, we suspect that high-frequency relative phase noise between the \SI{866}{\nano\meter} and \SI{397}{\nano\meter} lasers results in a deterioration of the dark state, further restricting the experimentally observed cooling rate at high Rabi frequencies.

In conclusion, we developed double-bright EIT (D-EIT) cooling as a novel scalable approach to standard EIT cooling and experimentally demonstrated D-EIT GSC of all three motional degrees of freedom of a single $^{40}$Ca$^+$ ion within one cooling pulse. We developed a time-dependent rate theory of the system beyond the Lamb-Dicke regime and show a quantitative match for the steady-state mean motional quantum number and qualitative agreement for the dependence of the cooling rate on the detuning. In the $^{40}$Ca$^+$ level structure, cooling rates for single EIT cooling are limited by off-resonant scattering via the unused magnetic sublevel of the $P_\mathrm{1/2}$ state. The chosen level scheme of D-EIT protects the dark state from such decoherence processes and thus, for systems with smaller Lamb-Dicke factors, it can reach higher cooling rates and lower $n_\mathrm{ss}$ and outperform single EIT cooling \cite{Scharnhorst_PRA}. 

The D-EIT technique is attractive for e.g. multi-ion, multi-species experiments in which quantum control of several spectrally separated motional modes is required. Examples include ion strings to study many-body physics \cite{schneider_experimental_2012, Kiethe_Probing_2017}, quantum simulations \cite{porras_effective_2004, blatt_quantum_2012}, and quantum information processing \cite{kienzler_observation_2016}. The technique will be particularly useful for quantum logic spectroscopy \cite{ wan_precision_2014, wolf_non-destructive_2016, rosenband_frrequency_2008, Schmidt2005, hempel_entanglement-enhanced_2013, chou_preparation_2017}, and quantum logic clocks \cite{rosenband_frrequency_2008, Ludlow_ClocksReview} for which two ions of a different species need to be cooled to suppress systematic frequency shifts. Notably, in a quantum logic clock with $^{27}$Al$^+$ as clock ion and a $^{40}$Ca$^+$ as logic ion, it is possible to choose a set of experimental parameters, where all six motional modes of the two ion crystal bunch at two frequencies. Here, the D-EIT scheme enables ground state cooling of all six modes of the two-ion crystal in one single cooling pulse. Furthermore, the D-EIT cooling scheme can provide even triple EIT cooling if the magnetic field strength is chosen such that the $|^2D_{3/2}, m_\mathrm{J}=\mp1/2\rangle$ sub-level is red detuned from the virtual excited state $P_+^v$ by the frequency of the radial mode in the case of \SI{397}{\nano\meter} being $\sigma^\pm$ polarized. That way, the D-EIT scheme additionally provides a second dark state at the position of the radial motional blue sideband, suppressing all second order radial heating processes \cite{evers_double-eit_2004}. In principle, it is possible to generate additional bright states for additional GSC around other frequencies by extending the D-EIT scheme through coupling more ground states from the $^2D_{3/2}$ manifold to the virtual excited states.

\begin{acknowledgments}
	The authors would like to thank Nicolas Spethmann for helpful comments on the manuscript. We acknowledge support from the DFG through CRC 1128 (geo-Q), project A03 and CRC 1227 (DQ-mat), project B0, and the state of Lower Saxony, Hannover, Germany. 
\end{acknowledgments}

\bibliography{Main}

\end{document}